\documentclass{ws-procs9x6}

\begin{document}

\title{Cross sections, error bars and event distributions in simulated 
Drell-Yan azimuthal asymmetry measurements.  
}

\author{A.~BIANCONI 
}

\address{Dip. Chimica e Fisica per l'ingegneria e i materiali, \\
Via Valotti 9, \\ 
Brescia 25100, Italy\\ 
E-mail: bianconi@bs.infn.it}



\maketitle

\abstracts{
A short summary of results of recent simulations of 
(un)polarized Drell-Yan experiments is presented here. Dilepton 
production in 
$pp$, $\bar{p}p$, $\pi^-p$ and $\pi^+p$ scattering is considered, 
for several kinematics corresponding to interesting 
regions for experiments at GSI, CERN-Compass and RHIC. 
A table of integrated cross sections, 
and a set of estimated error bars on measurements of 
azimuthal asymmetries (associated with collection of 5, 20 or 80 Kevents) 
are reported. 
}

\section{Introduction}

The aim of this work is to give some useful reference numbers for 
planning Drell-Yan experiments aimed at the measurement of 
transverse spin/momentum related azimuthal asymmetries. 

\begin{table}	
\tbl{Total cross sections (nb) for several colliding particle 
combinations, mass ranges and S-values. 
\label{table1}}
{\begin{tabular}{@{}ccccccccc@{}}
\hline
\multicolumn{9}{c}{}\\[-2ex]
M (GeV/c$^2$) &{} &1.5-2.5 &4-9 &12-40 &{} &1.5-2.5 &4-9 &12-40
\\
\hline
\multicolumn{9}{c}{}\\[-2ex]
S (GeV$^2$)
&{} &{} 
&$pp$  
&{} &{} &{}
&$\bar{p}p$ 
&{} 
\\
\hline
\multicolumn{9}{c}{}\\[-2ex]
30 &{} 
      &0.03 &$<<$pb &$<<$pb                &{}    &1.3 &0.3 pb &$<<$pb\\ 
200 
      &{} &0.7 &0.01 &$<<$pb               &{}    &4.4   &0.35 &$<<$pb\\
(200)$^2$ 
      &{} &3.5-17 &1.2-5.5 &0.06-0.26      &{}    &5-18 &2.6-7 &0.24-0.4\\
\hline
{} &{} &{} 
&$\pi^-p$ 
&{} &{} &{}
&$\pi^+p$ 
&{} 
\\
\hline
\multicolumn{9}{c}{}\\[-2ex]
30 
     &{} &0.9 &1 pb &$<<$pb             &{}   &0.25 &0.1 pb &$<<$pb\\ 
200
     &{} &1.9 &0.25 &$<<$pb             &{}   &0.7 &0.07 &$<<$pb\\
(200)$^2$ 
     &{} &1.8-5.6 &0.75-2.1 &0.1-0.2    &{}   &1.5-4.8 &0.5-1.4 &0.04-0.1\\
\hline
\end{tabular}}
\begin{tabnote}
For the high-energy case, two different parameterizations have been 
used, leading to pairs of $\sigma$ values. See text for details. 
\end{tabnote}
\end{table}

\begin{figure}[ht]
\centerline{\epsfxsize=4.1in\epsfbox{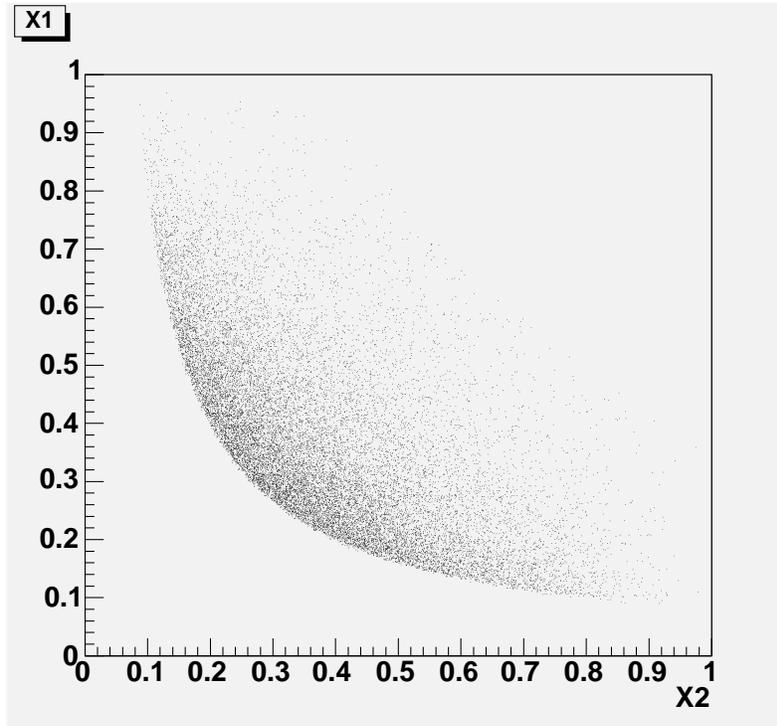}}   
\caption{$\bar{p}p$ event scatter plot for $S$ $=$ 200 GeV$^2$, and 
dilepton mass in the range 4-9 GeV/c$^2$. 
\label{scp1}}
\end{figure}

\begin{figure}[ht]
\centerline{\epsfxsize=4.1in\epsfbox{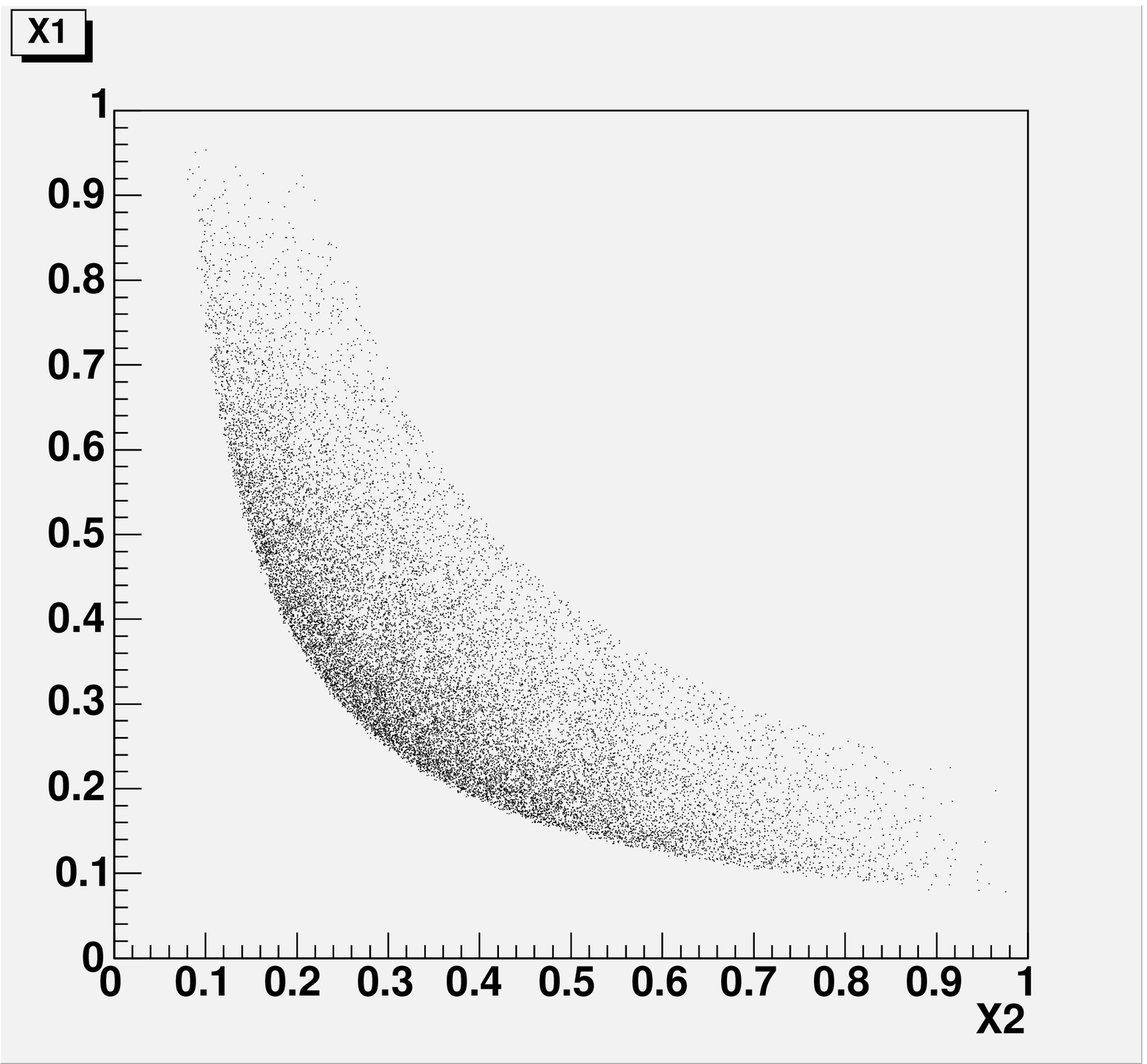}}   
\caption{$\bar{p}p$ event scatter plot for $S$ $=$ 30 GeV$^2$, and 
dilepton mass in the range 1.5-2.5 GeV/c$^2$. 
\label{scp2}}
\end{figure}

\begin{figure}[ht]
\centerline{\epsfxsize=4.1in\epsfbox{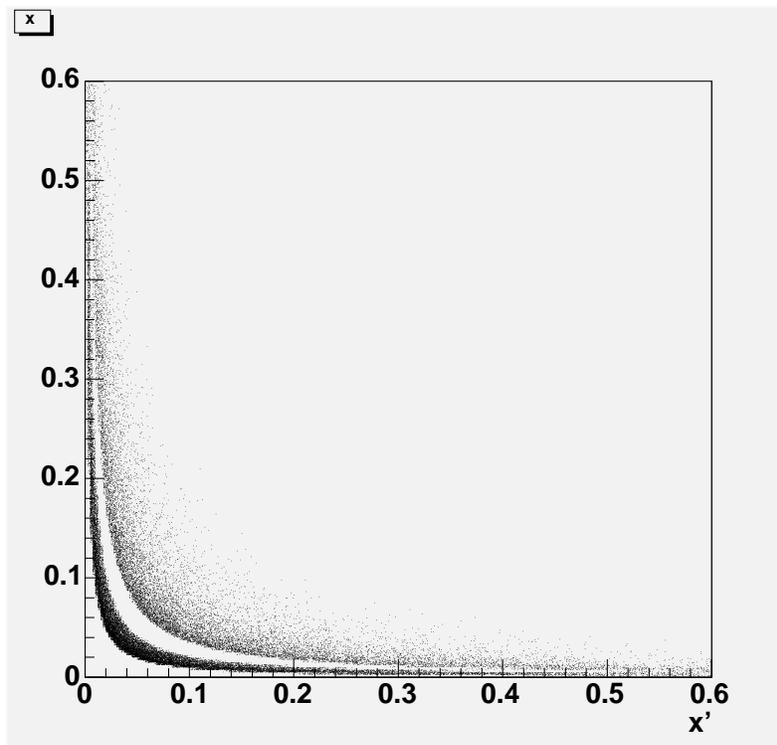}}   
\caption{$pp$ event scatter plot for $S$ $=$ (200)$^2$ GeV$^2$, 
and dilepton mass in the double range 4-9 and 12-40 GeV/c$^2$. 
\label{scp3}}
\end{figure}

\begin{figure}[ht]
\epsfxsize=5cm   
\centerline{\epsfxsize=4.1in\epsfbox{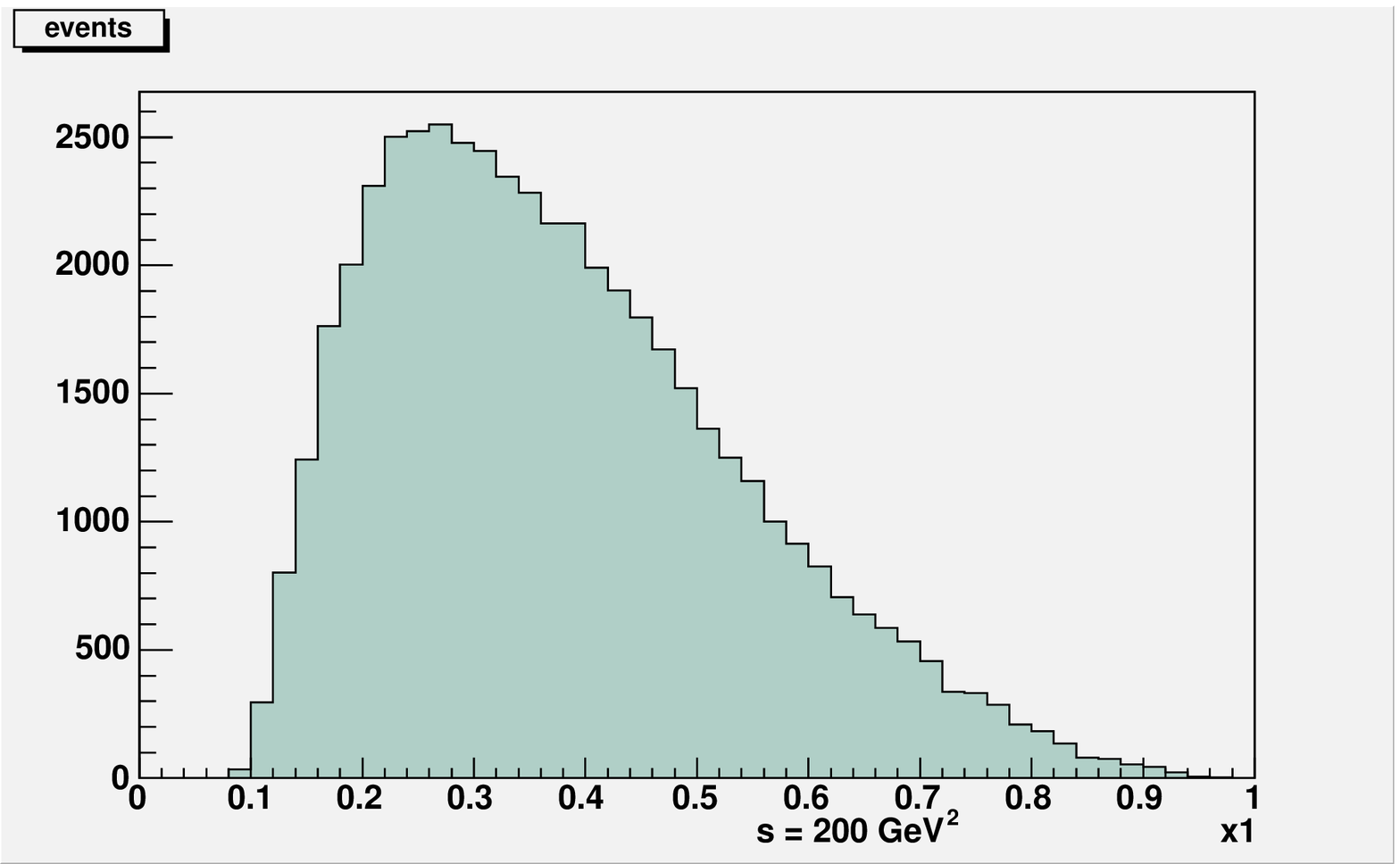}}   
\caption{X2-integrated distribution of the scatter plot of fig.1. 
Events are divided into 50 X1-bins. 
\label{hysto}}
\end{figure}

This includes total cross sections for several 
kinematical options for $pp$, $\bar{p}p$ and $\pi_\pm p$ Drell-Yan 
dilepton production: 
squared hadron-CM 
energy $S$ $=$ 
30 GeV$^2$, 200 GeV$^2$, (200)$^2$ GeV$^2$, and dilepton masses in  
the ranges 1.5-2.5 GeV/c$^2$, 4-9 GeV/c$^2$, 12-40 GeV/c$^2$. 
For some relevant situations, estimates of the asymmetry 
error bars are reported, 
for sets of 5, 20, 80 Kevents divided into 10 bins 
of the longitudinal fraction $x$. 
Previous Drell-Yan data and fitting 
relations\cite{Conway,Anass} are the basis 
of the initial core of the used simulation code.  
Several details on the formalism, 
together with the most recent examples of 
simulated asymmetries, are presented elsewhere in this workshop\cite{MR_here}, 
and in published work by myself and 
M.Radici\cite{BR_I,BR_II}. 

\section{Total Cross Sections.}

Total cross sections are shown in table 1. For 
the two lower $S-$values they 
have been evaluated with the differential cross-section 
fit relations\cite{Conway,Anass} coming from 
measurements of $\pi^-A$ and $\bar{p}A$ at $S$ $=$ 250-400 GeV$^2$. 
The two cases $\pi^-p$ and $\bar{p}p$ correspond to these cross sections 
for $Z/A$ $=$ 1. 
For $\pi^+p$ we have assumed 
$\pi^-p$ $\equiv$ $\pi^+p$ for the pion sea contribution, 
$\pi^-p$ $\equiv$ (1/4)$\pi^+n$ for the pion valence contribution, 
and calculated $\pi^-A$ for $Z/A$ $=$ 0. Cross sections for 
$pp$ have been evaluated by substituting 
the $(sea+valence)(sea+valence)$ structure of $\bar{p}p$ 
with the structure 
${1/2}[sea(sea+valence) + (sea+valence)sea]$. 
For the largest $S$ case the calculation based on the previous 
parameterization (with sea $\sim$ const for 
small $x$) has been sided by an alternative calculation 
using the (LO-intermediate gluon) MRST distributions \cite{Martin}, 
with sea $\sim$ $x^{-\lambda}$ 
($\lambda$ $\approx$ 0.2-0.3 for mass $<$ 9 GeV/c), and based 
on data sets including several recent Drell-Yan 
meaasurements\cite{DY_new}. No evolution was applied, which 
is unproper for $M$ $>>$ 10 GeV/c$^2$. 
The $K-$factors have been assumed as constant 
but tuned to reproduce with both methods the measured cross sections  
at $S$ $=$ 250 GeV$^2$, masses 4-9 GeV/c$^2$\cite{Anass}. 
The former strategy leads to the smaller reported cross section 
values, the latter to the bigger ones. 
For the $\pi^\pm p$ high-energy case 
double values refer to different parameterizations for the 
distribution functions of the proton only, pion distribution 
functions have not been changed. 
For the lowest mass range 1.5-2.5 the smaller number of contributing 
quarks introduces a reduction factor $\approx$ 1/2. 
The difference between the other two 
mass ranges is not essential. 

\section{Event distributions}

The event distribution has the general form 

\begin{equation}
N(S, x, x',P_T, \xi)\ =\  F(S,x,x',P_T)\cdot [1\ +\ A(x,x',P_T,\xi)]. 
\end{equation}
\noindent 
where $x,x',P_T$ describe the virtual photon 
kinematics in the 
hadron center of mass 
($P_L/(S/2)$ $\equiv$ $x-x'$,  
$M^2/S$ $\equiv$ $xx'$), 
while $\xi$ represents compactly 
the set of variables describing the angular distribution 
of the leptons in the Collins-Soper frame ($\equiv$ 
the photon polarization). 
$F(S,x,x',P_T)$ alone gives the virtual photon event distribution. 
$A(x,x',P_T,\xi)$ averages to zero over all the solid angle, and 
describes the asymmetry properties of the lepton distribution 
in unpolarized, single or double polarized DY. 

The scatter plots of figs 1,2 and 3 reproduce the 
event distribution 
\begin{equation} 
N(S,x,x')\ =\ \int d^2 \vec P_T F(S,x,x',P_T)
\end{equation}
for some relevant $\bar{p}p$ and $pp$ cases. 
Fig.4 reports
\begin{equation}
N(S,x)\ =\ \int dx' N(S,x,x')
\end{equation}
where the integrated distribution is the one of fig.1. 
Events in the $xx'$ scatter plots concentrate near the 
hyperbole $xx'$ $=$ ${M^2}_{min}/S$, because $\sigma$ $\propto$ 
$q(x)\bar{q}(x')/M^2$. 
For the same reason, in the 
case of fig.3 the lower event band ($M$ in the range 4-9 GeV/c$^2$) 
contains 95 \% of all the events reported in the figure. 

\section{Asymmetry Error Bars}

\begin{figure}[ht]
\centerline{\epsfxsize=4.1in\epsfbox{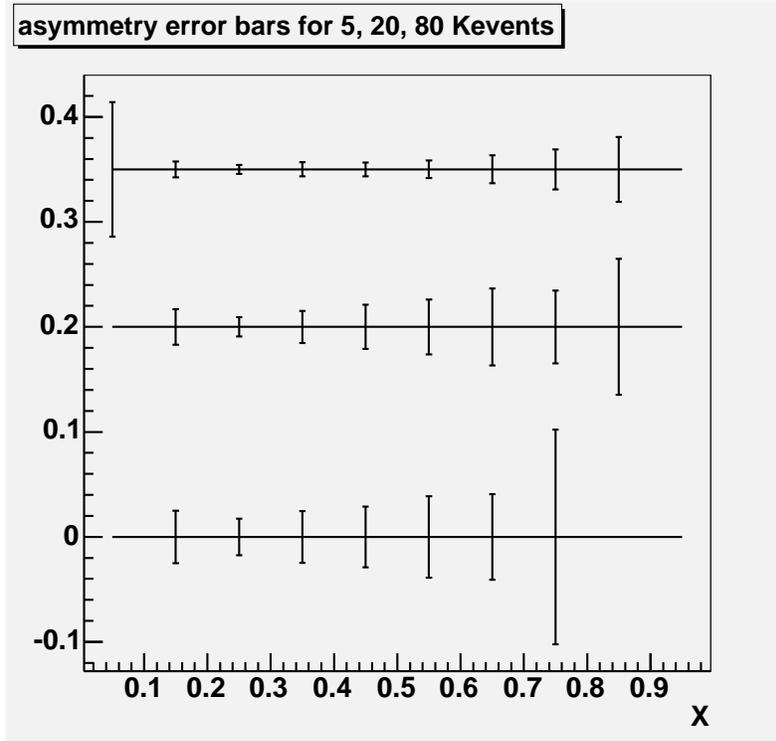}}   
\caption{Error bars on the azimuthal asymmetry 
for $\bar{p}p$, $S$ $=$ 
30 GeV$^2$, dilepton mass in the range 1.5-2.5 GeV/c$^2$. 
\label{error1}}
\end{figure}

\begin{figure}[ht]
\centerline{\epsfxsize=4.1in\epsfbox{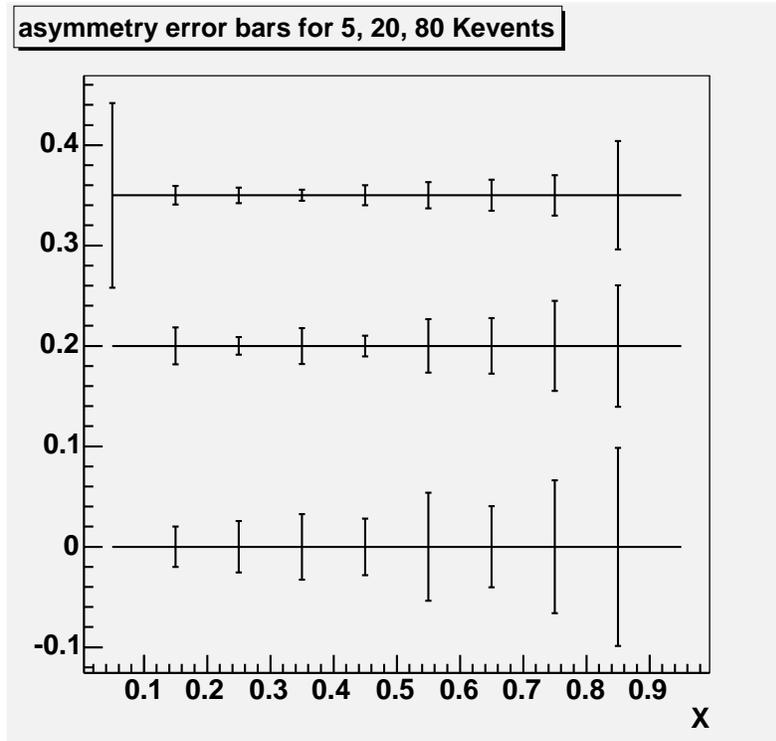}}   
\caption{Error bars on the azimuthal asymmetry 
for $\pi^-p$, pion beam energy 100 GeV, fixed 
target, dilepton mass in the range 4-9 GeV/c$^2$. 
\label{error2}}
\end{figure}

\begin{figure}[ht]
\centerline{\epsfxsize=4.1in\epsfbox{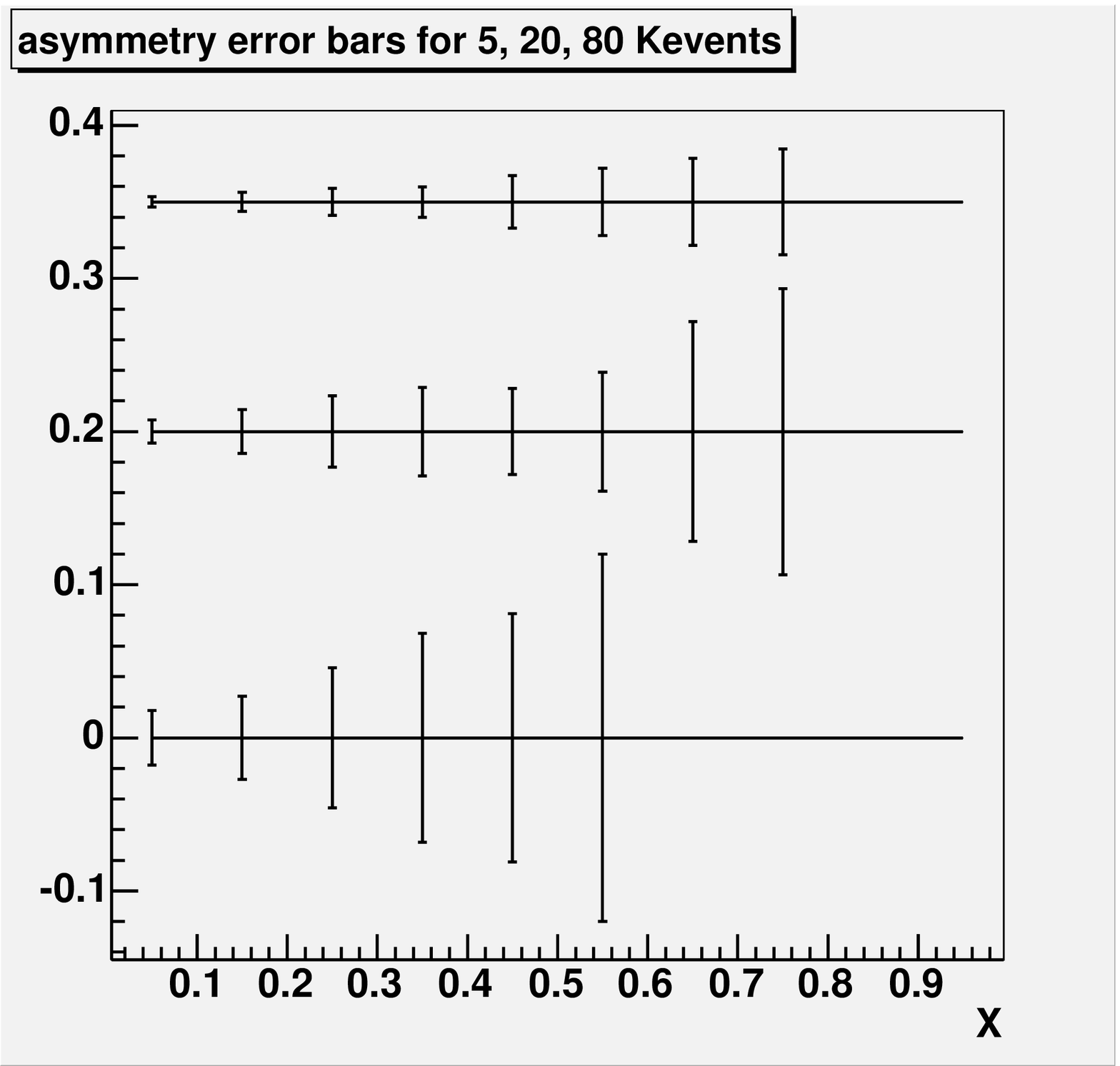}}   
\caption{Error bars on the azimuthal asymmetry 
for $pp$, $S$ $=$ (200)$^2$ GeV$^2$, dilepton 
mass in the range 12-40 GeV/c$^2$. 
\label{error3}}
\end{figure}

The integrated $N(S,x)$ distribution has its peak at $x$ 
$\approx$ $3{M^2}_{min}/S$. To the right of the peak, 
$N(S,x)$ $\sim$ $1/x^n$ with $n$ $>$ 1. This selects 
for each $S$, $M_{min}$ the $x-$range where error bars are smaller. 
The shown error bars in figs.5, 6 and 7 
refer to the special case of Sivers asymmetry, 
however they are the same for any kind of left/right asymmetry with 
respect to a Collins-Soper azimuthal angle $\phi$. The asymmetry is 
defined as $(A-B)/(A+B)$, where $A$ and $B$ are the event numbers 
with positive or negative $sin(\phi-\phi_S)$.  
Error bars have been 
calculated by assuming constant 0.05 asymmetry everywhere, and 
repeating the simulation 10 times to calculate fluctuations. 
Despite the size of the error bars is 
reasonably stable for $N_{repetitions}$ $>$ 5, 
with $N_{repetitions}$ $=$ 10 there are still 
small but evident fluctuations in the error bar size. This is due to 
most $x-$bins being filled with event numbers 
$N(+)+N(-)$ $<$ 1000. 
For $N(+)+N(-)$ $<$ 50 error bars are 
not shown. Typically the largest shown error bars refer to event numbers 
$N(+)+N(-)$ $\sim$ 100. 
Examination of the error bar fluctuations 
suggests that the reported error bars are 
reliable within a factor $1/\sqrt{2}\div\sqrt{2}$. 
If one repeats the calculation by assuming 
asymmetry 0 or 0.1, systematic changes of these (purely 
statistic) error bars are smaller than the above fluctuations.  
So, unless they damp asymmetries by orders, all those 
reducing coefficient 
like polarization dilution etc are not influent on the error bar 
sizes. For asymmetry $>$ 33 \%, 
$N(-)$ $<$ $N(+)/2$. So, in the case of really large 
asymmetries, the consequent small value of $N(-)$ 
in those bins whose population is $<<$ 1000 
introduces errors whose size may be 
uncontrollably larger than the estimated ones. In this case 
the error simulation must be more specific to be appropriate.


\begin{thebibliography}{0}

\bibitem{Conway} J.S.Conway et al, {\it Phys. Rev.} {\bf D39}, 
92 (1989).

\bibitem{Anass}  E.Anassontzis et al, 
{\it Phys. Rev.} {\bf D38}, 1377 (1988).

\bibitem{MR_here}  M.Radici, this workshop.

\bibitem{BR_I} A.Bianconi and M.Radici, {\it Phys. Rev.}
{\bf D34}, 1729 (1980).

\bibitem{BR_II} A.Bianconi and M.Radici {\it Phys. Rev.}  
{\bf D25}, L527 (1992).

\bibitem{Martin}  A.D.Martin et al, {\it Eur. Phys. J.} 
{\bf C4} (1998) 463, and {\it Phys. Lett.} {\bf B443} (1998) 301.

\bibitem{DY_new}  E605 collaboration, G.Moreno et al, {\it Phys. Rev.} 
{\bf D43} (1991) 2815, E772 collaboration, P.L.McGaughey et al, 
{\it Phys. Rev.} {\bf D50} (1994) 3038, NA51 collaboration, A.Baldit et 
al, {\it Phys. Lett.} {\bf B332} (1994) 244.
E866 collaboration, E.A.Hawker et al,  
{\it Phys. Rev. Lett.} 
{\bf 80} (1998) 3715
 




\end{thebibliography}
\end{document}